\documentclass{article}
\usepackage{spconf,amsmath,amssymb,graphicx,booktabs,url}
\usepackage{tikz}
\usepackage[T1]{fontenc}
\usepackage[utf8]{inputenc}
\usepackage[hidelinks]{hyperref}

\usepackage{siunitx}
\usetikzlibrary{arrows.meta,positioning,fit,backgrounds,shapes.geometric,calc}

\graphicspath{{gen/}{figs/}{./}}

\usepackage{etoolbox}
\makeatletter
\long\def\@makecaption#1#2{%
 \vskip 10pt
 \setbox\@tempboxa\hbox{#1. #2}%
 \ifdim \wd\@tempboxa >\hsize #1. #2\par \else \hbox
to\hsize{\hfil\box\@tempboxa\hfil}%
 \fi
 \vskip 6pt}
\makeatother

\newif\ifsubmission
\submissiontrue

\title{Retrieving Individual Stems from Music Mixtures with Slot Embeddings}

\name{David Braun$^{\star}$, \enspace Junyi Fan$^{\circ}$,
\enspace Pranay Manocha$^{\diamond}$, \enspace Donald S.\ Williamson$^{\circ}$,
\enspace Adam Finkelstein$^{\star}$}

\address{
$^{\star}$ Princeton University \qquad
$^{\circ}$ The Ohio State University \qquad
$^{\diamond}$ Symbal AI \\
{\normalsize\mbox{
  \texttt{\{db1224,af\}@princeton.edu}\enspace
  \texttt{\{fan.1188,williamson.413\}@osu.edu}\enspace
  \texttt{pranay@symbal.ai}}}}

\begin{document}
\ninept
\maketitle

\newcommand{\SongsTrain}{166}
\newcommand{\TrainStems}{1485}
\newcommand{\TrainArtistGroups}{20}
\newcommand{\SongsValidation}{37}
\newcommand{\ValidationStems}{282}
\newcommand{\ValidationArtistGroups}{13}
\newcommand{\SongsTest}{35}
\newcommand{\TestStems}{301}
\newcommand{\TestArtistGroups}{9}
\newcommand{\SongsTotal}{238}
\newcommand{\DbSize}{2068}
\newcommand{\SlakhSongs}{150}
\newcommand{\SlakhDbSize}{1458}
\newcommand{\TrainingSeeds}{3}
\newcommand{\EvaluationRounds}{1}
\newcommand{\MoisesEvalTargets}{4608}
\newcommand{\MoisesEvalMixtures}{1536}
\newcommand{\MoisesEvalVisits}{384}
\newcommand{\SlakhEvalTargets}{5400}
\newcommand{\SlakhEvalMixtures}{1800}
\newcommand{\SlakhEvalVisits}{450}
\newcommand{\PredictedCountSlots}{2.99}
\newcommand{\PredictedCountSlotsSd}{0.06}
\newcommand{\PredictedCountAccuracy}{69.9}
\newcommand{\PredictedCountAccuracySd}{1.5}
\newcommand{\PredictedCountZeroSlots}{0.00}
\newcommand{\PredictedCountZeroSlotsSd}{0.00}
\newcommand{\SlakhPredictedCountSlots}{2.85}
\newcommand{\SlakhPredictedCountSlotsSd}{0.12}
\newcommand{\SlakhPredictedCountAccuracy}{49.8}
\newcommand{\SlakhPredictedCountAccuracySd}{4.8}
\newcommand{\SlakhPredictedCountZeroSlots}{0.59}
\newcommand{\SlakhPredictedCountZeroSlotsSd}{0.28}
\newcommand{\PairedStembedMinusSameSongNegativesOneMean}{48.9}
\newcommand{\PairedSameSongNegativesMinusStembedOneMean}{-48.9}
\newcommand{\PairedStembedMinusSameSongNegativesOneSd}{0.5}
\newcommand{\PairedSameSongNegativesMinusStembedOneSd}{0.5}
\newcommand{\PairedStembedMinusSameSongNegativesFamilyOneMean}{49.9}
\newcommand{\PairedSameSongNegativesMinusStembedFamilyOneMean}{-49.9}
\newcommand{\PairedStembedMinusSameSongNegativesFamilyOneSd}{1.0}
\newcommand{\PairedSameSongNegativesMinusStembedFamilyOneSd}{1.0}
\newcommand{\PairedStembedMinusSameSongNegativesFiveMean}{59.5}
\newcommand{\PairedSameSongNegativesMinusStembedFiveMean}{-59.5}
\newcommand{\PairedStembedMinusSameSongNegativesFiveSd}{0.8}
\newcommand{\PairedSameSongNegativesMinusStembedFiveSd}{0.8}
\newcommand{\PairedStembedMinusSameSongNegativesFamilyFiveMean}{59.3}
\newcommand{\PairedSameSongNegativesMinusStembedFamilyFiveMean}{-59.3}
\newcommand{\PairedStembedMinusSameSongNegativesFamilyFiveSd}{1.1}
\newcommand{\PairedSameSongNegativesMinusStembedFamilyFiveSd}{1.1}
\newcommand{\PairedStembedMinusCrossSongNegativesOneMean}{11.5}
\newcommand{\PairedCrossSongNegativesMinusStembedOneMean}{-11.5}
\newcommand{\PairedStembedMinusCrossSongNegativesOneSd}{2.1}
\newcommand{\PairedCrossSongNegativesMinusStembedOneSd}{2.1}
\newcommand{\PairedStembedMinusCrossSongNegativesFamilyOneMean}{8.8}
\newcommand{\PairedCrossSongNegativesMinusStembedFamilyOneMean}{-8.8}
\newcommand{\PairedStembedMinusCrossSongNegativesFamilyOneSd}{1.7}
\newcommand{\PairedCrossSongNegativesMinusStembedFamilyOneSd}{1.7}
\newcommand{\PairedStembedMinusCrossSongNegativesFiveMean}{7.8}
\newcommand{\PairedCrossSongNegativesMinusStembedFiveMean}{-7.8}
\newcommand{\PairedStembedMinusCrossSongNegativesFiveSd}{1.7}
\newcommand{\PairedCrossSongNegativesMinusStembedFiveSd}{1.7}
\newcommand{\PairedStembedMinusCrossSongNegativesFamilyFiveMean}{3.5}
\newcommand{\PairedCrossSongNegativesMinusStembedFamilyFiveMean}{-3.5}
\newcommand{\PairedStembedMinusCrossSongNegativesFamilyFiveSd}{1.7}
\newcommand{\PairedCrossSongNegativesMinusStembedFamilyFiveSd}{1.7}
\newcommand{\PairedStembedMinusSharedWindowOneMean}{-0.2}
\newcommand{\PairedSharedWindowMinusStembedOneMean}{0.2}
\newcommand{\PairedStembedMinusSharedWindowOneSd}{1.5}
\newcommand{\PairedSharedWindowMinusStembedOneSd}{1.5}
\newcommand{\PairedStembedMinusSharedWindowFamilyOneMean}{0.0}
\newcommand{\PairedSharedWindowMinusStembedFamilyOneMean}{-0.0}
\newcommand{\PairedStembedMinusSharedWindowFamilyOneSd}{2.0}
\newcommand{\PairedSharedWindowMinusStembedFamilyOneSd}{2.0}
\newcommand{\PairedStembedMinusSharedWindowFiveMean}{3.2}
\newcommand{\PairedSharedWindowMinusStembedFiveMean}{-3.2}
\newcommand{\PairedStembedMinusSharedWindowFiveSd}{0.9}
\newcommand{\PairedSharedWindowMinusStembedFiveSd}{0.9}
\newcommand{\PairedStembedMinusSharedWindowFamilyFiveMean}{4.0}
\newcommand{\PairedSharedWindowMinusStembedFamilyFiveMean}{-4.0}
\newcommand{\PairedStembedMinusSharedWindowFamilyFiveSd}{0.9}
\newcommand{\PairedSharedWindowMinusStembedFamilyFiveSd}{0.9}
\newcommand{\PairedStembedMinusCrossSongMixturesOneMean}{-0.6}
\newcommand{\PairedCrossSongMixturesMinusStembedOneMean}{0.6}
\newcommand{\PairedStembedMinusCrossSongMixturesOneSd}{1.8}
\newcommand{\PairedCrossSongMixturesMinusStembedOneSd}{1.8}
\newcommand{\PairedStembedMinusCrossSongMixturesFamilyOneMean}{-0.5}
\newcommand{\PairedCrossSongMixturesMinusStembedFamilyOneMean}{0.5}
\newcommand{\PairedStembedMinusCrossSongMixturesFamilyOneSd}{2.2}
\newcommand{\PairedCrossSongMixturesMinusStembedFamilyOneSd}{2.2}
\newcommand{\PairedStembedMinusCrossSongMixturesFiveMean}{1.5}
\newcommand{\PairedCrossSongMixturesMinusStembedFiveMean}{-1.5}
\newcommand{\PairedStembedMinusCrossSongMixturesFiveSd}{0.2}
\newcommand{\PairedCrossSongMixturesMinusStembedFiveSd}{0.2}
\newcommand{\PairedStembedMinusCrossSongMixturesFamilyFiveMean}{2.3}
\newcommand{\PairedCrossSongMixturesMinusStembedFamilyFiveMean}{-2.3}
\newcommand{\PairedStembedMinusCrossSongMixturesFamilyFiveSd}{0.2}
\newcommand{\PairedCrossSongMixturesMinusStembedFamilyFiveSd}{0.2}
\newcommand{\PairedStembedMinusSlotTripletOneMean}{-1.1}
\newcommand{\PairedSlotTripletMinusStembedOneMean}{1.1}
\newcommand{\PairedStembedMinusSlotTripletOneSd}{1.0}
\newcommand{\PairedSlotTripletMinusStembedOneSd}{1.0}
\newcommand{\PairedStembedMinusSlotTripletFamilyOneMean}{-1.2}
\newcommand{\PairedSlotTripletMinusStembedFamilyOneMean}{1.2}
\newcommand{\PairedStembedMinusSlotTripletFamilyOneSd}{1.3}
\newcommand{\PairedSlotTripletMinusStembedFamilyOneSd}{1.3}
\newcommand{\PairedStembedMinusSlotTripletFiveMean}{-1.0}
\newcommand{\PairedSlotTripletMinusStembedFiveMean}{1.0}
\newcommand{\PairedStembedMinusSlotTripletFiveSd}{0.5}
\newcommand{\PairedSlotTripletMinusStembedFiveSd}{0.5}
\newcommand{\PairedStembedMinusSlotTripletFamilyFiveMean}{-0.7}
\newcommand{\PairedSlotTripletMinusStembedFamilyFiveMean}{0.7}
\newcommand{\PairedStembedMinusSlotTripletFamilyFiveSd}{0.5}
\newcommand{\PairedSlotTripletMinusStembedFamilyFiveSd}{0.5}
\newcommand{\PairedStembedMinusPredictedCountOneMean}{0.9}
\newcommand{\PairedPredictedCountMinusStembedOneMean}{-0.9}
\newcommand{\PairedStembedMinusPredictedCountOneSd}{0.4}
\newcommand{\PairedPredictedCountMinusStembedOneSd}{0.4}
\newcommand{\PairedStembedMinusPredictedCountFamilyOneMean}{1.1}
\newcommand{\PairedPredictedCountMinusStembedFamilyOneMean}{-1.1}
\newcommand{\PairedStembedMinusPredictedCountFamilyOneSd}{0.4}
\newcommand{\PairedPredictedCountMinusStembedFamilyOneSd}{0.4}
\newcommand{\PairedStembedMinusPredictedCountFiveMean}{1.8}
\newcommand{\PairedPredictedCountMinusStembedFiveMean}{-1.8}
\newcommand{\PairedStembedMinusPredictedCountFiveSd}{0.6}
\newcommand{\PairedPredictedCountMinusStembedFiveSd}{0.6}
\newcommand{\PairedStembedMinusPredictedCountFamilyFiveMean}{2.5}
\newcommand{\PairedPredictedCountMinusStembedFamilyFiveMean}{-2.5}
\newcommand{\PairedStembedMinusPredictedCountFamilyFiveSd}{0.6}
\newcommand{\PairedPredictedCountMinusStembedFamilyFiveSd}{0.6}
\newcommand{\PairedStembedMinusCirMuqOneMean}{33.2}
\newcommand{\PairedCirMuqMinusStembedOneMean}{-33.2}
\newcommand{\PairedStembedMinusCirMuqOneSd}{0.5}
\newcommand{\PairedCirMuqMinusStembedOneSd}{0.5}
\newcommand{\PairedStembedMinusCirMuqFamilyOneMean}{6.6}
\newcommand{\PairedCirMuqMinusStembedFamilyOneMean}{-6.6}
\newcommand{\PairedStembedMinusCirMuqFamilyOneSd}{2.0}
\newcommand{\PairedCirMuqMinusStembedFamilyOneSd}{2.0}
\newcommand{\PairedStembedMinusCirMuqFiveMean}{17.4}
\newcommand{\PairedCirMuqMinusStembedFiveMean}{-17.4}
\newcommand{\PairedStembedMinusCirMuqFiveSd}{0.6}
\newcommand{\PairedCirMuqMinusStembedFiveSd}{0.6}
\newcommand{\PairedStembedMinusCirMuqFamilyFiveMean}{-0.5}
\newcommand{\PairedCirMuqMinusStembedFamilyFiveMean}{0.5}
\newcommand{\PairedStembedMinusCirMuqFamilyFiveSd}{1.0}
\newcommand{\PairedCirMuqMinusStembedFamilyFiveSd}{1.0}
\newcommand{\PairedStembedMinusSmtiOneMean}{35.4}
\newcommand{\PairedSmtiMinusStembedOneMean}{-35.4}
\newcommand{\PairedStembedMinusSmtiOneSd}{1.4}
\newcommand{\PairedSmtiMinusStembedOneSd}{1.4}
\newcommand{\PairedStembedMinusSmtiFamilyOneMean}{35.3}
\newcommand{\PairedSmtiMinusStembedFamilyOneMean}{-35.3}
\newcommand{\PairedStembedMinusSmtiFamilyOneSd}{1.8}
\newcommand{\PairedSmtiMinusStembedFamilyOneSd}{1.8}
\newcommand{\PairedStembedMinusSmtiFiveMean}{38.4}
\newcommand{\PairedSmtiMinusStembedFiveMean}{-38.4}
\newcommand{\PairedStembedMinusSmtiFiveSd}{0.7}
\newcommand{\PairedSmtiMinusStembedFiveSd}{0.7}
\newcommand{\PairedStembedMinusSmtiFamilyFiveMean}{37.2}
\newcommand{\PairedSmtiMinusStembedFamilyFiveMean}{-37.2}
\newcommand{\PairedStembedMinusSmtiFamilyFiveSd}{1.0}
\newcommand{\PairedSmtiMinusStembedFamilyFiveSd}{1.0}
\newcommand{\StembedParametersM}{191.44}
\newcommand{\StembedBackboneParametersM}{122.94}
\newcommand{\StembedParametersRoundedM}{191}
\newcommand{\StembedBackboneParametersRoundedM}{123}
\newcommand{\StembedWarmupSteps}{125}
\newcommand{\StembedReadout}{4}
\newcommand{\StembedBatchInputs}{320}
\newcommand{\StembedTrainingUpdates}{3000}
\newcommand{\StembedWidth}{1024}
\newcommand{\StembedEmbeddingDim}{256}
\newcommand{\StembedDecoderBlocks}{4}
\newcommand{\StembedEncoderBlocks}{0}
\newcommand{\StembedSlots}{4}
\newcommand{\StembedEmbeddingHidden}{0}
\newcommand{\StembedPresenceHidden}{1024}
\newcommand{\CirMuqParametersM}{219.64}
\newcommand{\CirMuqBackboneParametersM}{219.64}
\newcommand{\CirMuqParametersRoundedM}{220}
\newcommand{\CirMuqBackboneParametersRoundedM}{220}
\newcommand{\CirMuqWarmupSteps}{125}
\newcommand{\CirMuqReadout}{8}
\newcommand{\CirMuqBatchInputs}{320}
\newcommand{\CirMuqTrainingUpdates}{3000}
\newcommand{\FilteredStembedOne}{60.9}
\newcommand{\FilteredStembedOneSd}{0.8}
\newcommand{\FilteredStembedFive}{79.6}
\newcommand{\FilteredStembedFiveSd}{0.1}
\newcommand{\FilteredStembedFamilyOne}{63.4}
\newcommand{\FilteredStembedFamilyOneSd}{1.2}
\newcommand{\FilteredStembedFamilyFive}{82.6}
\newcommand{\FilteredStembedFamilyFiveSd}{0.2}
\newcommand{\FilteredGalleryMin}{1995}
\newcommand{\FilteredGalleryMinSd}{0}
\newcommand{\FilteredGalleryMax}{2068}
\newcommand{\FilteredGalleryMaxSd}{0}
\newcommand{\FilteredGalleryMean}{2031}
\newcommand{\FilteredGalleryMeanSd}{0}
\newcommand{\FilteredCirMuqOne}{25.4}
\newcommand{\FilteredCirMuqOneSd}{0.7}
\newcommand{\FilteredCirMuqFive}{63.8}
\newcommand{\FilteredCirMuqFiveSd}{0.9}
\newcommand{\FilteredCirMuqFamilyOne}{56.6}
\newcommand{\FilteredCirMuqFamilyOneSd}{1.0}
\newcommand{\FilteredCirMuqFamilyFive}{83.2}
\newcommand{\FilteredCirMuqFamilyFiveSd}{1.3}
\newcommand{\CountTwoAccuracy}{61.9}
\newcommand{\CountTwoAccuracySd}{3.9}
\newcommand{\CountTwoMae}{0.42}
\newcommand{\CountTwoMaeSd}{0.05}
\newcommand{\CountTwoUnder}{7.2}
\newcommand{\CountTwoUnderSd}{0.9}
\newcommand{\CountTwoOver}{30.8}
\newcommand{\CountTwoOverSd}{4.4}
\newcommand{\CountTwoMixtures}{2304}
\newcommand{\CountTwoMixturesSd}{0}
\newcommand{\CountThreeAccuracy}{70.8}
\newcommand{\CountThreeAccuracySd}{1.2}
\newcommand{\CountThreeMae}{0.30}
\newcommand{\CountThreeMaeSd}{0.01}
\newcommand{\CountThreeUnder}{15.4}
\newcommand{\CountThreeUnderSd}{2.5}
\newcommand{\CountThreeOver}{13.8}
\newcommand{\CountThreeOverSd}{3.8}
\newcommand{\CountThreeMixtures}{1536}
\newcommand{\CountThreeMixturesSd}{0}
\newcommand{\CountFourAccuracy}{43.3}
\newcommand{\CountFourAccuracySd}{2.0}
\newcommand{\CountFourMae}{0.61}
\newcommand{\CountFourMaeSd}{0.01}
\newcommand{\CountFourUnder}{56.7}
\newcommand{\CountFourUnderSd}{2.0}
\newcommand{\CountFourOver}{0.0}
\newcommand{\CountFourOverSd}{0.0}
\newcommand{\CountFourMixtures}{384}
\newcommand{\CountFourMixturesSd}{0}
\newcommand{\SmtiTrainingUpdates}{3000}
\newcommand{\SmtiTrainingUpdatesSd}{0}
\newcommand{\SmtiEmbeddingDim}{1024}
\newcommand{\SmtiEmbeddingDimSd}{0}
\newcommand{\SmtiTeacherInputs}{512}
\newcommand{\SmtiTeacherInputsSd}{0}
\newcommand{\SmtiMixtureInputs}{192}
\newcommand{\SmtiMixtureInputsSd}{0}
\newcommand{\FilteredSmtiOne}{22.5}
\newcommand{\FilteredSmtiOneSd}{0.5}
\newcommand{\FilteredSmtiFive}{40.7}
\newcommand{\FilteredSmtiFiveSd}{0.8}
\newcommand{\FilteredSmtiFamilyOne}{26.1}
\newcommand{\FilteredSmtiFamilyOneSd}{0.4}
\newcommand{\FilteredSmtiFamilyFive}{45.5}
\newcommand{\FilteredSmtiFamilyFiveSd}{1.0}
\newcommand{\FilteredPredictedCountOne}{59.9}
\newcommand{\FilteredPredictedCountOneSd}{1.1}
\newcommand{\FilteredPredictedCountFamilyOne}{62.1}
\newcommand{\FilteredPredictedCountFamilyOneSd}{1.6}
\newcommand{\FilteredPredictedCountFive}{77.7}
\newcommand{\FilteredPredictedCountFiveSd}{0.7}
\newcommand{\FilteredPredictedCountFamilyFive}{80.0}
\newcommand{\FilteredPredictedCountFamilyFiveSd}{0.7}
\newcommand{\TripletMargin}{0.40}
\newcommand{\TripletMarginSd}{0.00}
\newcommand{\TripletWeight}{16.00}
\newcommand{\TripletWeightSd}{0.00}
\newcommand{\OverlappingTargetQueries}{24}
\newcommand{\OverlappingTargetQueriesSd}{0}
\newcommand{\OverlappingTargetPercent}{0.52}
\newcommand{\OverlappingTargetPercentSd}{0.00}
\newcommand{\StembedOne}{56.8}
\newcommand{\StembedOneSd}{0.9}
\newcommand{\StembedFamilyOne}{58.9}
\newcommand{\StembedFamilyOneSd}{1.3}
\newcommand{\StembedFive}{78.1}
\newcommand{\StembedFiveSd}{0.2}
\newcommand{\StembedFamilyFive}{81.2}
\newcommand{\StembedFamilyFiveSd}{0.1}
\newcommand{\PredictedCountOne}{56.0}
\newcommand{\PredictedCountOneSd}{1.3}
\newcommand{\PredictedCountFamilyOne}{57.8}
\newcommand{\PredictedCountFamilyOneSd}{1.7}
\newcommand{\PredictedCountFive}{76.3}
\newcommand{\PredictedCountFiveSd}{0.5}
\newcommand{\PredictedCountFamilyFive}{78.7}
\newcommand{\PredictedCountFamilyFiveSd}{0.6}
\newcommand{\GtStembedOne}{75.1}
\newcommand{\GtStembedOneSd}{0.7}
\newcommand{\GtStembedFamilyOne}{75.5}
\newcommand{\GtStembedFamilyOneSd}{0.7}
\newcommand{\GtStembedFive}{89.1}
\newcommand{\GtStembedFiveSd}{0.8}
\newcommand{\GtStembedFamilyFive}{89.6}
\newcommand{\GtStembedFamilyFiveSd}{0.4}
\newcommand{\DemucsStembedOne}{53.2}
\newcommand{\DemucsStembedOneSd}{0.5}
\newcommand{\DemucsStembedFamilyOne}{55.9}
\newcommand{\DemucsStembedFamilyOneSd}{1.1}
\newcommand{\DemucsStembedFive}{72.3}
\newcommand{\DemucsStembedFiveSd}{0.6}
\newcommand{\DemucsStembedFamilyFive}{75.8}
\newcommand{\DemucsStembedFamilyFiveSd}{0.7}
\newcommand{\CirMuqOne}{23.6}
\newcommand{\CirMuqOneSd}{0.5}
\newcommand{\CirMuqFamilyOne}{52.4}
\newcommand{\CirMuqFamilyOneSd}{0.8}
\newcommand{\CirMuqFive}{60.7}
\newcommand{\CirMuqFiveSd}{0.6}
\newcommand{\CirMuqFamilyFive}{81.7}
\newcommand{\CirMuqFamilyFiveSd}{1.1}
\newcommand{\CrossSongNegativesOne}{45.3}
\newcommand{\CrossSongNegativesOneSd}{1.3}
\newcommand{\CrossSongNegativesFamilyOne}{50.1}
\newcommand{\CrossSongNegativesFamilyOneSd}{0.4}
\newcommand{\CrossSongNegativesFive}{70.3}
\newcommand{\CrossSongNegativesFiveSd}{1.9}
\newcommand{\CrossSongNegativesFamilyFive}{77.8}
\newcommand{\CrossSongNegativesFamilyFiveSd}{1.6}
\newcommand{\CrossSongMixturesOne}{57.5}
\newcommand{\CrossSongMixturesOneSd}{0.9}
\newcommand{\CrossSongMixturesFamilyOne}{59.4}
\newcommand{\CrossSongMixturesFamilyOneSd}{0.9}
\newcommand{\CrossSongMixturesFive}{76.7}
\newcommand{\CrossSongMixturesFiveSd}{0.3}
\newcommand{\CrossSongMixturesFamilyFive}{78.9}
\newcommand{\CrossSongMixturesFamilyFiveSd}{0.2}
\newcommand{\SameSongNegativesOne}{7.9}
\newcommand{\SameSongNegativesOneSd}{0.4}
\newcommand{\SameSongNegativesFamilyOne}{9.0}
\newcommand{\SameSongNegativesFamilyOneSd}{0.7}
\newcommand{\SameSongNegativesFive}{18.6}
\newcommand{\SameSongNegativesFiveSd}{0.8}
\newcommand{\SameSongNegativesFamilyFive}{21.9}
\newcommand{\SameSongNegativesFamilyFiveSd}{1.0}
\newcommand{\SharedWindowOne}{57.0}
\newcommand{\SharedWindowOneSd}{0.7}
\newcommand{\SharedWindowFamilyOne}{58.9}
\newcommand{\SharedWindowFamilyOneSd}{1.0}
\newcommand{\SharedWindowFive}{75.0}
\newcommand{\SharedWindowFiveSd}{0.9}
\newcommand{\SharedWindowFamilyFive}{77.3}
\newcommand{\SharedWindowFamilyFiveSd}{0.9}
\newcommand{\SlotTripletOne}{58.0}
\newcommand{\SlotTripletOneSd}{0.5}
\newcommand{\SlotTripletFamilyOne}{60.1}
\newcommand{\SlotTripletFamilyOneSd}{0.5}
\newcommand{\SlotTripletFive}{79.2}
\newcommand{\SlotTripletFiveSd}{0.5}
\newcommand{\SlotTripletFamilyFive}{81.9}
\newcommand{\SlotTripletFamilyFiveSd}{0.4}
\newcommand{\CirOne}{10.7}
\newcommand{\CirFamilyOne}{20.1}
\newcommand{\CirFive}{22.3}
\newcommand{\CirFamilyFive}{41.3}
\newcommand{\GtCirOne}{64.8}
\newcommand{\GtCirFamilyOne}{65.6}
\newcommand{\GtCirFive}{81.3}
\newcommand{\GtCirFamilyFive}{81.7}
\newcommand{\DemucsCirOne}{34.5}
\newcommand{\DemucsCirFamilyOne}{36.0}
\newcommand{\DemucsCirFive}{55.0}
\newcommand{\DemucsCirFamilyFive}{57.5}
\newcommand{\MuqOne}{10.7}
\newcommand{\MuqFamilyOne}{16.0}
\newcommand{\MuqFive}{17.1}
\newcommand{\MuqFamilyFive}{28.1}
\newcommand{\MuqMulanOne}{6.0}
\newcommand{\MuqMulanFamilyOne}{12.1}
\newcommand{\MuqMulanFive}{13.6}
\newcommand{\MuqMulanFamilyFive}{26.6}
\newcommand{\FigmaOne}{10.1}
\newcommand{\FigmaFamilyOne}{20.0}
\newcommand{\FigmaFive}{20.7}
\newcommand{\FigmaFamilyFive}{39.4}
\newcommand{\MertOne}{5.4}
\newcommand{\MertFamilyOne}{9.4}
\newcommand{\MertFive}{10.8}
\newcommand{\MertFamilyFive}{20.2}
\newcommand{\ClapOne}{4.3}
\newcommand{\ClapFamilyOne}{6.7}
\newcommand{\ClapFive}{9.7}
\newcommand{\ClapFamilyFive}{18.6}
\newcommand{\SmtiOne}{21.4}
\newcommand{\SmtiOneSd}{0.6}
\newcommand{\SmtiFamilyOne}{23.7}
\newcommand{\SmtiFamilyOneSd}{0.5}
\newcommand{\SmtiFive}{39.8}
\newcommand{\SmtiFiveSd}{0.9}
\newcommand{\SmtiFamilyFive}{44.0}
\newcommand{\SmtiFamilyFiveSd}{0.9}
\newcommand{\GtFrozenMuqOne}{52.8}
\newcommand{\GtFrozenMuqFamilyOne}{53.4}
\newcommand{\GtFrozenMuqFive}{65.1}
\newcommand{\GtFrozenMuqFamilyFive}{66.9}
\newcommand{\SlakhStembedOne}{38.1}
\newcommand{\SlakhStembedOneSd}{1.1}
\newcommand{\SlakhStembedFamilyOne}{43.2}
\newcommand{\SlakhStembedFamilyOneSd}{1.1}
\newcommand{\SlakhStembedFive}{59.0}
\newcommand{\SlakhStembedFiveSd}{1.5}
\newcommand{\SlakhStembedFamilyFive}{68.0}
\newcommand{\SlakhStembedFamilyFiveSd}{1.2}
\newcommand{\SlakhPredictedCountOne}{37.5}
\newcommand{\SlakhPredictedCountOneSd}{0.1}
\newcommand{\SlakhPredictedCountFamilyOne}{41.4}
\newcommand{\SlakhPredictedCountFamilyOneSd}{0.4}
\newcommand{\SlakhPredictedCountFive}{56.7}
\newcommand{\SlakhPredictedCountFiveSd}{0.7}
\newcommand{\SlakhPredictedCountFamilyFive}{62.9}
\newcommand{\SlakhPredictedCountFamilyFiveSd}{0.4}
\newcommand{\SlakhCirMuqOne}{20.2}
\newcommand{\SlakhCirMuqOneSd}{0.3}
\newcommand{\SlakhCirMuqFamilyOne}{42.6}
\newcommand{\SlakhCirMuqFamilyOneSd}{0.5}
\newcommand{\SlakhCirMuqFive}{45.1}
\newcommand{\SlakhCirMuqFiveSd}{0.3}
\newcommand{\SlakhCirMuqFamilyFive}{70.0}
\newcommand{\SlakhCirMuqFamilyFiveSd}{0.7}
\newcommand{\SlakhSmtiOne}{15.8}
\newcommand{\SlakhSmtiOneSd}{0.7}
\newcommand{\SlakhSmtiFamilyOne}{19.3}
\newcommand{\SlakhSmtiFamilyOneSd}{0.5}
\newcommand{\SlakhSmtiFive}{29.4}
\newcommand{\SlakhSmtiFiveSd}{0.5}
\newcommand{\SlakhSmtiFamilyFive}{36.8}
\newcommand{\SlakhSmtiFamilyFiveSd}{0.6}

\begin{abstract}
Music producers search libraries of isolated instrument recordings, called stems, for sounds resembling parts of an existing song.
Neural retrieval systems address this by mapping audio to embeddings and ranking library stems by their similarity to the query.
The leading method, Contrastive Instrument Retrieval (CIR), encodes the mixture as a single embedding, but it works best when a user specifies the target's instrument family.
We introduce Stembed, which encodes a mixture as several slot embeddings representing candidate stems.
During training, we construct mixtures from stems of the same song and match their slot embeddings to those of the isolated stems.
The slot embeddings from mixtures inherit the stem identities of their assigned solo embedding, enabling a contrastive loss.
On mixtures from held out MoisesDB artists, Stembed outperforms a CIR-style baseline when both search the full stem library.
Even when predicting the stem count itself without family labels, Stembed exceeds the baseline's family-filtered R@1.
Our website demonstrates how users can select a slot by inspecting the tags of its retrieved stems.
\end{abstract}


\section{Introduction}

Musicians often use existing recordings to communicate the sounds they want in a production.
A reference recording may contain a desirable guitar, drum kit, or synthesizer, each of which could guide a search through a stem library.
Audio retrieval models encode the recording as a single embedding~\cite{vaillant_contrastive_2026} or a set of embeddings~\cite{kim_show_2023}, then compare these with embeddings of library stems.
A single mixture embedding aggregates information from all its underlying stems.
We study how to encode the mixture as a set of embeddings, each providing a potential query to a library.

One approach is to apply source separation~\cite{rouard_hybrid_2023} and encode the resulting stems for retrieval~\cite{vaillant_contrastive_2026}.
However, separation artifacts can affect the embeddings, and predefined output categories can group several instruments together.
Methods that retrieve directly from mixtures offer another approach.
Show Me the Instruments (SMTI)~\cite{kim_show_2023} first trains an instrument encoder through classification, then trains a mixture encoder to predict a set of embeddings from the frozen instrument encoder.
The second stage uses permutation-invariant training with minimum-cost matching~\cite{kuhn1955hungarian,dovrat_many_2021}.

More recently, Contrastive Instrument Retrieval (CIR)~\cite{vaillant_contrastive_2026} fine-tunes an Audio Spectrogram Transformer (AST)~\cite{gong_ast_2021} trained on AudioSet~\cite{gemmeke_audio_2017} to encode both mixtures and isolated instruments.
Its contrastive objective brings a mixture's embedding close to those of its stems.
Since multiple instruments compete for representation in the same coordinate space, CIR benefits from searching within a specified instrument family.
However, specifying a family requires skilled human intervention and limits automation.

Slot-based models~\cite{carion_detr_2020, locatello_object_2020} provide another way to produce a set of latent embeddings.
A slot is a learned representation intended to describe an individual component of the input.
AudioSlots~\cite{reddy_audioslots_2023} maps two-speaker mixtures to source slots, which guide the reconstruction of spectrograms of the individual speakers.
MusicSlots~\cite{gha_unsupervised_2023} learns slots for individual notes in synthesized chords through unsupervised spectrogram reconstruction.
Compositional Audio Representation Learning (CARL)~\cite{sridhar_compositional_2025} learns source-centric slots from frozen audio features using classification supervision or feature reconstruction.

We introduce Stembed, which learns stem-specific slot embeddings from multitrack recordings.
Unlike SMTI, whose second training stage predicts frozen solo embeddings, Stembed fine-tunes a joint representation of stems and mixtures.
During training, we match a mixture's slots to solo embeddings.
The matched slots inherit the corresponding stem identities, enabling a contrastive loss.
The matching procedure leaves some slots assigned and some unassigned.
A slot predicts its own binary assignment with a ``presence'' score.
At inference, this score enables selecting which slot embeddings become queries.
In principle, users could inspect each slot's retrieved tags or filenames and choose which results to audition without naming an instrument family.
This could turn the \textit{production} task of specifying a family into a \textit{recognition} task.

\textbf{Our main contributions} are as follows.
First, we introduce a framework for contrastive training of stem-level slots.
Second, we evaluate the approach on real multitrack music, comparing against CIR- and SMTI-style objectives with a consistent audio backbone.
We also compare against pipelines that first source separate and then encode.
On MoisesDB, without instrument-family guidance, Stembed achieves higher R@1 than family-guided CIR--MuQ.
Code, model weights, and a retrieval demo are available.\footnote{\url{http://dbraun.github.io/stembed}}
The demo suggests that slots specialize in instrument types within mixtures, with slot 1 tending to represent vocals when they're present.

\suppressfloats[t]
\begin{figure}[t]
\centering
\resizebox{0.9\linewidth}{!}{
\begin{tikzpicture}[
  font=\small,
  yscale=0.8,
  >={Latex[length=1.6mm]},
  enc/.style={draw,trapezium,shape border rotate=270,fill=black!6,
              trapezium stretches=true,font=\scriptsize,minimum width=4.8mm,minimum height=4.8mm,inner sep=0.4pt},
  pred/.style={draw,trapezium,shape border rotate=270,fill=black!6,
               trapezium stretches=true,font=\scriptsize,minimum width=4.8mm,minimum height=4.8mm,inner sep=0.4pt},
  dec/.style={draw,rounded corners=1pt,fill=black!6,font=\scriptsize,minimum width=4mm,minimum height=5.2mm,inner sep=0.4pt},
  slot/.style={draw,rounded corners=1pt,font=\scriptsize,minimum width=8mm,minimum height=4mm,inner sep=0.4pt},
  barz/.style={draw,rounded corners=1pt,minimum width=10mm,minimum height=5.4mm,
               inner sep=0.5pt,densely dashed},
  one/.style={draw,rounded corners=1pt,minimum width=8mm,minimum height=5.3mm,
              inner sep=0.8pt,fill=#1},
  two/.style 2 args={draw,rounded corners=1pt,minimum width=8mm,minimum height=5.3mm,inner sep=0.8pt,
    path picture={
      \fill[#1] (path picture bounding box.north west) rectangle (path picture bounding box.east);
      \fill[#2] (path picture bounding box.west) rectangle (path picture bounding box.south east);
    }},
  three/.style n args={3}{draw,rounded corners=1pt,minimum width=8mm,minimum height=5.3mm,inner sep=0.8pt,
    path picture={
      \fill[#1] (path picture bounding box.north west) rectangle (path picture bounding box.south east);
      \fill[#2] ($(path picture bounding box.north west)!1/3!(path picture bounding box.south west)$) rectangle (path picture bounding box.south east);
      \fill[#3] ($(path picture bounding box.north west)!2/3!(path picture bounding box.south west)$) rectangle (path picture bounding box.south east);
    }},
  tied/.style={densely dotted},
  sg/.style={dashed,->},
  loss/.style={draw,rounded corners=1pt,inner sep=1.5pt,fill=white},
  lbl/.style={font=\scriptsize\itshape,text=black!85},
  hdr/.style={font=\scriptsize\bfseries},
]
\definecolor{okBlue}{HTML}{0072B2}
\definecolor{okOrange}{HTML}{D55E00}
\definecolor{okGreen}{HTML}{009E73}
\definecolor{okPurple}{HTML}{CC79A7}
\colorlet{cA}{okBlue}\colorlet{cB}{okOrange}
\colorlet{cC}{okGreen}\colorlet{cD}{okPurple}
\colorlet{fA}{okBlue!18}\colorlet{fB}{okOrange!20}
\colorlet{fC}{okGreen!18}\colorlet{fD}{okPurple!24}
\colorlet{cE}{black!60}\colorlet{cF}{black!85}
\colorlet{fE}{black!8}\colorlet{fF}{black!20}


\node[one={fA}]      (nA)  at (0,0.55)  {\fontsize{6}{7}\selectfont\textcolor{cA}{$A$}};
\node[two={fA}{fB}]  (nAB) at (0,-0.85) {\fontsize{6}{7}\selectfont\textcolor{cA}{$A$}\,\textcolor{cB}{$B$}};
\node[three={fA}{fB}{fC}] (nABC) at (0,-2.48) {\fontsize{6}{7}\selectfont\textcolor{cA}{$A$}\,\textcolor{cB}{$B$}\,\textcolor{cC}{$C$}};
\node at (0,-3.30) {$\vdots$};
\node[two={fE}{fF}]  (nEF) at (0,-3.95) {\fontsize{6}{7}\selectfont\textcolor{cE}{$A'$}\,\textcolor{cF}{$B'$}};

\foreach \n/\y in {nA/0.55, nAB/-0.85, nABC/-2.48, nEF/-3.95}{
  \node[enc] (E\n) at (1.25,\y) {$\mathcal{E}$};
  \node[dec] (D\n) at (2.2,\y) {$\mathcal{D}$};
  \draw[->] (\n) -- (E\n);
  \draw[->] (E\n) -- (D\n);
}
\draw[tied] (EnA) -- (EnAB) -- (EnABC);
\draw[tied] (DnA) -- (DnAB) -- (DnABC);
\draw[tied] (EnABC) -- (EnEF);
\draw[tied] (DnABC) -- (DnEF);

\node[slot,fill=fA] (sA)   at (3.55,0.55)  {$z^{A}_{A}$};
\node[slot,fill=fA] (sABa) at (3.55,-0.53) {$z^{AB}_{A}$};
\node[slot,fill=fB] (sABb) at (3.55,-1.17) {$z^{AB}_{B}$};
\node[slot,fill=fA] (sABCa) at (3.55,-1.92) {$z^{ABC}_{A}$};
\node[slot,fill=fB] (sABCb) at (3.55,-2.48) {$z^{ABC}_{B}$};
\node[slot,fill=fC] (sABCc) at (3.55,-3.04) {$z^{ABC}_{C}$};
\node[slot,fill=fE] (sEFa) at (3.55,-3.63) {$z^{A'B'}_{A'}$};
\node[slot,fill=fF] (sEFb) at (3.55,-4.27) {$z^{A'B'}_{B'}$};
\draw[->] (DnA)  -- (sA);
\foreach \e/\a/\b in {DnAB/sABa/sABb, DnEF/sEFa/sEFb}{
  \draw[->] (\e.east) to[out=0,in=180] (\a.west);
  \draw[->] (\e.east) to[out=0,in=180] (\b.west);
}
\foreach \slotname in {sABCa,sABCb,sABCc}{
  \draw[->] (DnABC.east) to[out=0,in=180] (\slotname.west);
}
\node[lbl,align=left,anchor=north west] at (nA.west |- 0,-4.8) {$n{=}4$: 4 solos; 2 pairs: $AB, CD$;\\4 triples: $ABC, ABD, ACD, BCD$; 10 nodes};

\draw[<->,semithick,cA] (sA.east) .. controls +(1.05,-0.1) and +(1.05,0.35) ..
  node[pos=0.5,right,inner sep=2pt]{\scriptsize share $A$} (sABa.east);
\draw[<->,semithick,cA] (sABa.east) .. controls +(1.05,-0.1) and +(1.05,0.35) ..
  node[pos=0.5,right,inner sep=2pt]{\scriptsize share $A$} (sABCa.east);
\draw[<->,densely dashed,black!55] (sABCa.east) .. controls +(1.05,-0.1) and +(1.05,0.35) ..
  node[pos=0.5,right,inner sep=2pt,align=left]{\scriptsize same-song\\[-1pt]\scriptsize negative} (sABCc.east);
\draw[<->,densely dashed,black!55] (sABCc.east) .. controls +(1.05,-0.1) and +(1.05,0.35) ..
  node[pos=0.5,right,inner sep=2pt,align=left]{\scriptsize cross-song\\[-1pt]\scriptsize negative} (sEFb.east);

\end{tikzpicture}%
}
\caption{
Training graph for $n{=}4$ stems, showing $A$, $AB$, $ABC$, and another song's $A'B'$.
$\mathcal{E}$: backbone; $\mathcal{D}$: slot decoder.
$z_S^X$ is node $X$'s slot matched to stem $S$; only matched slots are shown.
Dotted links denote shared weights, and solid links denote positives for $A$.
}
\label{fig:graph}
\end{figure}

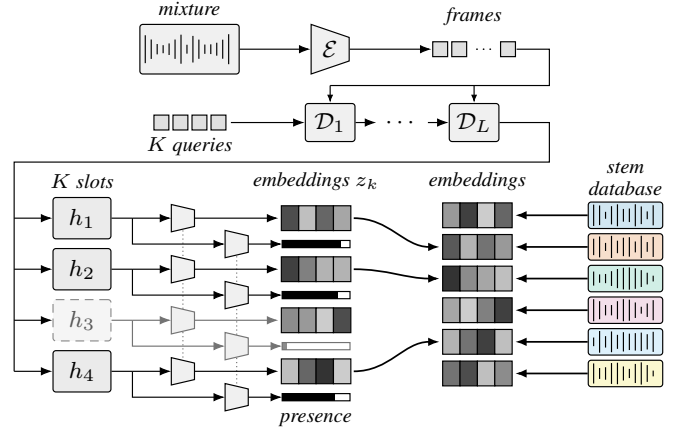
\begin{figure}[t]
\centering
\resizebox{1.0\linewidth}{!}{
\begin{tikzpicture}[
 yscale=.87,
 font=\fontsize{8}{9}\selectfont,text=black,
 >={Latex[length=1.4mm]},line width=.4pt,
 enc/.style={draw,trapezium,shape border rotate=270,fill=black!6,trapezium stretches=true,minimum width=6mm,minimum height=4.6mm,inner sep=.5pt},
 dec/.style={draw,rounded corners=1pt,fill=black!6,minimum width=6mm,minimum height=4.6mm,inner sep=.5pt},
 slot/.style={draw,rounded corners=1pt,minimum width=7mm,minimum height=5mm,inner sep=.5pt,fill=black!8},
 slotoff/.style={slot,densely dashed,draw=black!45,fill=black!4,text=black!65},
 stem/.style={draw,rounded corners=1pt,fill=black!6,minimum width=9mm,minimum height=3.2mm,inner sep=.5pt},
 lbl/.style={font=\fontsize{7}{8}\selectfont\itshape,inner sep=.5pt},
 wf/.style={line width=.4pt}]
\definecolor{okBlue}{HTML}{0072B2}
\definecolor{okOrange}{HTML}{D55E00}
\definecolor{okGreen}{HTML}{009E73}
\definecolor{okPurple}{HTML}{CC79A7}
\definecolor{okSkyBlue}{HTML}{56B4E9}
\definecolor{okYellow}{HTML}{F0E442}
\colorlet{stemFill1}{okBlue!18}
\colorlet{stemFill2}{okOrange!20}
\colorlet{stemFill3}{okGreen!18}
\colorlet{stemFill4}{okPurple!24}
\colorlet{stemFill5}{okSkyBlue!20}
\colorlet{stemFill6}{okYellow!25}
\begin{scope}[xshift=12mm]
\begin{scope}[yshift=1.5mm]
\node[draw,rounded corners=1pt,fill=black!6,minimum width=12mm,minimum height=6mm] (mix) at (1.1,0) {};
\foreach \j in {1,...,15}{
 \draw[wf] ($(mix.west)+(0.072*\j,{-0.05-0.17*abs(sin(\j*67))*abs(cos(\j*23))})$)
 -- ($(mix.west)+(0.072*\j,{0.05+0.17*abs(sin(\j*67))*abs(cos(\j*23))})$);}
\node[lbl] at (1.1,.56) {mixture};
\node[enc] (E) at (2.8,0) {$\mathcal E$};
\foreach \i/\x in {1/4.13,2/4.36,3/4.95}{
 \node[draw,fill=black!12,minimum width=1.8mm,minimum height=1.8mm,inner sep=0pt] (f\i) at (\x,0) {};}
\coordinate (dots) at ($(f2.east)!.5!(f3.west)$);
\foreach \dx in {-.08,0,.08}{\fill ($(dots)+(\dx,0)$) circle[radius=.11mm];}
\node[fit=(f1)(f3),inner sep=0pt] (frames) {};
\node[lbl] at (4.54,.46) {frames};
\draw[->] (mix) -- (E);
\draw[->] (E) -- (frames);
\end{scope}
\node[dec] (D1) at (2.8,-.87) {$\mathcal D_1$};
\node[dec] (D2) at (4.54,-.87) {$\mathcal D_L$};
\foreach \i/\x in {1/.76,2/.99,3/1.22,4/1.45}{
 \node[draw,fill=black!12,minimum width=1.8mm,minimum height=1.8mm,inner sep=0pt] (q\i) at (\x,-.87) {};}
\node[lbl] at (1.1,-1.2) {$K$ queries};
\draw[->] (1.6,-.87) -- (D1.west);
\node (decoderDots) at ($(D1.east)!.5!(D2.west)$) {$\cdots$};
\draw[->] (D1.east) -- (decoderDots.west);
\draw[->] (decoderDots.east) -- (D2.west);
\coordinate (frameBus) at ($(E.south)!.5!(D1.north)$);
\draw (frames.east) -- (5.45,0 |- frames.east)
 -- (5.45,0 |- frameBus) -- (D1.north |- frameBus);
\draw[-{Latex[length=1mm]}] (D1.north |- frameBus) -- (D1.north);
\draw[-{Latex[length=1mm]}] (D2.north |- frameBus) -- (D2.north);
\end{scope}
\draw (D2.east) -- (6.65,-.87) -- (6.65,-1.42) -- (.18,-1.42) -- (.18,-4.33);
\node[lbl] at (1.0,-1.68) {$K$ slots};
\node[lbl] at (3.83,-1.68) {embeddings $z_k$};
\node[lbl] at (5.78,-1.68) {embeddings};
\node[lbl,align=center,anchor=south] at (7.57,-1.91) {stem\\database};
\node[slot] (s1) at (1,-2.2) {$h_1$};
\node[slot] (s2) at (1,-2.91) {$h_2$};
\node[slotoff] (s3) at (1,-3.62) {$h_3$};
\node[slot] (s4) at (1,-4.33) {$h_4$};
\foreach \i in {1,...,4}{\draw[->] (.18,0 |- s\i.west) -- (s\i.west);}
\draw[densely dotted,black!55] (2.22,-2.2) -- (2.22,-4.33);
\draw[densely dotted,black!55] (2.87,-2.56) -- (2.87,-4.69);
\foreach \i/\y/\a/\b/\c/\d/\frac/\shade in {
 1/-2.2/70/25/60/35/.86/100,
 2/-2.91/75/50/30/30/.82/100,
 3/-3.62/45/40/20/70/.06/45,
 4/-4.33/25/60/80/20/.78/100}{
 \foreach \j/\v in {0/\a,1/\b,2/\c,3/\d}{
  \node[draw,fill=black!\v,minimum width=2.1mm,minimum height=3mm,inner sep=0pt] (q\i c\j) at (3.515+.21*\j,\y) {};}
 \node[fit=(q\i c0)(q\i c3),inner sep=.6pt] (qe\i) {};
 \ifnum\i=3\def\slotline{black!55}\else\def\slotline{black}\fi
 \draw[\slotline] (s\i.east) -- (1.61,\y);
 \foreach \dy/\dx in {0/0,-.36/.65}{
  \draw[draw=\slotline,fill=black!6]
   (2.08+\dx,\y+\dy-.19) -- (2.08+\dx,\y+\dy+.19)
   -- (2.36+\dx,\y+\dy+.115) -- (2.36+\dx,\y+\dy-.115) -- cycle;}
 \draw[->,draw=\slotline] (1.61,\y) -- (2.08,\y);
 \draw[->,draw=\slotline] (1.61,\y) |- (2.73,\y-.36);
 \draw[->,draw=\slotline] (2.36,\y) -- (qe\i.west);
 \draw[->,draw=\slotline] (3.01,\y-.36) -- (3.42,\y-.36);
 \fill[black!\shade] (3.42,\y-.41) rectangle ++(.82*\frac,.10);
 \draw[\slotline] (3.42,\y-.41) rectangle ++(.82,.10);}
\node[lbl] at (3.83,-5.0) {presence};
\foreach \i/\y/\a/\b/\c/\d in {
 1/-2.16/40/75/20/60,
 2/-2.60/65/30/55/40,
 3/-3.04/80/45/35/25,
 4/-3.48/35/20/50/75,
 5/-3.92/30/55/75/25,
 6/-4.36/55/70/25/45}{
 \foreach \j/\v in {0/\a,1/\b,2/\c,3/\d}{
  \node[draw,fill=black!\v,minimum width=2.1mm,minimum height=3mm,inner sep=0pt] (d\i c\j) at (5.465+.21*\j,\y) {};}
 \node[fit=(d\i c0)(d\i c3),inner sep=.6pt] (de\i) {};
 \node[stem,fill=stemFill\i] (db\i) at (7.57,\y) {};
 \draw[->,draw=black,semithick] (db\i.west) -- (de\i.east);
 \foreach \j in {1,...,11}{
 \draw[wf,black] ($(db\i.west)+(.072*\j,{-.04-.115*abs(sin(\j*53*\i+17*\i))})$)
 -- ($(db\i.west)+(.072*\j,{.04+.115*abs(sin(\j*53*\i+17*\i))})$);}}
\draw[->,black,semithick] (qe1.east) to[out=0,in=180] (de2.west);
\draw[->,black,semithick] (qe2.east) to[out=0,in=180] (de3.west);
\draw[->,black,semithick] (qe4.east) to[out=0,in=180] (de5.west);
\end{tikzpicture}%
}
\caption{
Inference.
$L$ decoder blocks refine $K{=}4$ learned queries by cross-attending to frame tokens from $\mathcal{E}$.
Shared heads map each slot $h_k$ to a presence logit and an embedding $z_k$.
With $m{=}3$, the three highest logits select $z_1$, $z_2$, and $z_4$ as queries.
}
\label{fig:deploy}
\end{figure}

\section{Method}
\label{sec:method}

Each training batch contains $B=32$ groups of $n{=}4$ stems, each sampled within one song.
From each group, we build ten audio nodes: four solos, two disjoint pairs, and all four triples (Fig.~\ref{fig:graph}), so each batch contains $32\times10=320$ audio inputs.
Each stem appears in five nodes: one solo, one pair, and three triples.
We call these nodes a graph because shared stems connect them.
Following~\cite{saeed_contrastive_2021} and~\cite[Sec.~3.1]{garoufis_multi-source_2023}, each node samples its own 5\,s window of the song and sums its stems.
Our strategy prefers to use non-overlapping windows for each node.
We normalize all nodes to \qty{-18}{\text{LUFS}} with a 24 dB gain cap and encode with the same network to place solo and mixture audio in one embedding space.

The decoder (\S\ref{ssec:decoder}) returns $K=4$ slots with corresponding embeddings $z_k\in\mathbb{R}^d$ and $k=1,\ldots,K$.
The \emph{solo path} encodes an isolated stem using only the first slot during training and database encoding.

Our objective combines contrastive learning over assigned slots with presence prediction.
For each $m$-stem training mixture, we assign the $m$ solo stem embeddings to slots, not allowing multiple solos to pick the same slot embedding.
The assignment minimizes total squared Euclidean distance between $\ell_2$-normalized embeddings, leaving $K-m$ slots unassigned.
Since $K$ and $m$ are small, we solve these assignments by exact enumeration rather than the Hungarian algorithm~\cite{kuhn1955hungarian} used by DETR~\cite{carion_detr_2020}.

In deployment, solo stems are unavailable to perform the matching procedure, meaning we don't know which slot embeddings are closest to the solo embeddings.
We therefore train a presence predictor using binary targets $y_{Xk}$ from the matching assignments: one for assigned slots and zero for unassigned slots.
The presence loss $\mathcal{L}_{\text{p}}$ averages binary cross-entropy over all slots of mixture nodes.
Let $\mathcal{X}_{\mathrm{mix}}$ be the batch's mixture nodes.
For slot $k$ of mixture $X$, let $a_{Xk}$ be the presence logit and $p_{Xk}=\sigma(a_{Xk})$.
Then
\begin{equation}
\begin{aligned}
\mathcal{L}_{\mathrm p}&=\frac{1}{K|\mathcal{X}_{\mathrm{mix}}|}
\sum_{X\in\mathcal{X}_{\mathrm{mix}}}\sum_{k=1}^{K}\ell(p_{Xk},y_{Xk}),\\
\ell(p,y)&=-y\log p-(1-y)\log(1-p).
\end{aligned}
\end{equation}

A solo embedding uses its stem's identity, and an assigned slot inherits the identity of its matched solo stem.
Two embeddings with the same identity form a positive pair, excluding self-pairs.
For example, in Fig.~\ref{fig:graph}, $z_A^A$, $z_A^{AB}$, and $z_A^{ABC}$ can form pairwise positives.
Embeddings of different stem identities form negative pairs.
Despite the discreteness of assignment, gradients flow through both solo and matched slot embeddings.

We collect solo and matched mixture embeddings from the batch into a list, $z_1,\ldots,z_M$.
Each graph contributes $4+2\times2+4\times3=20$ embeddings, giving $M=20B$ embeddings across the batch.
Unmatched slots do not enter this list.
Let $\mathcal{V}(i)$ be the other views of $z_i$'s stem and $s_{ij} = \cos(z_i, z_j)$.
Eq.~\eqref{eq:contrastive} is the supervised contrastive loss~\cite{khosla_supervised_2020} with stem identity as the label, a multi-positive form of InfoNCE~\cite{oord_representation_2018}.
With temperature $\tau{=}0.1$,
\begin{equation}
\label{eq:contrastive}
\mathcal{L}_{\text{c}} = -\frac{1}{M} \sum_{i=1}^{M}
  \frac{1}{|\mathcal{V}(i)|}
  \sum_{j \in \mathcal{V}(i)}
  \log \frac{\exp(s_{ij}/\tau)}
            {\sum_{\substack{q=1 \\ q\neq i}}^{M} \exp(s_{iq}/\tau)}.
\end{equation}
The training loss is $\mathcal{L} = \lambda_{\text{c}}\mathcal{L}_{\text{c}} + \lambda_{\text{p}} \mathcal{L}_{\text{p}}$, with $\lambda_{\text{c}}{=}1.0$ and $\lambda_{\text{p}}{=}0.1$.

\subsection{Transformer slot decoder}
\label{ssec:decoder}

A backbone $\mathcal{E}$ maps the waveform to a sequence of $T$ frame tokens of width $W=1024$ (Fig.~\ref{fig:deploy}).
Following~\cite{carion_detr_2020}, a transformer decoder~\cite{vaswani_attention_2017} refines $K=4$ learned query vectors of width $W$ through $L$ non-autoregressive blocks that cross-attend to the frame tokens.
Each block applies unmasked slot self-attention, cross-attention, and a feed-forward network, with a residual connection around each sublayer.
After layer normalization, separate output heads map each decoder state $h_k\in\mathbb{R}^W$ to a presence logit scalar and retrieval embedding $z_k\in\mathbb{R}^d$, with $k=1,\ldots,K$.

\noindent\textbf{Configuration.}
The backbone is MuQ~\cite{zhu_muq_2025}, a music self-supervised Conformer taking mono audio at \qty{24}{\kilo\hertz}, fine-tuned end-to-end.
The frame tokens carry temporal information from MuQ's rotary positional encoding, so the decoder adds no positional embeddings.
The selected model uses frame tokens after \StembedReadout{} MuQ Conformer blocks and uses \StembedDecoderBlocks{} pre-norm decoder blocks of width $W=\StembedWidth$, with four attention heads and a $4W=4096$-unit feed-forward layer per block.
The embedding head is linear with output dimension $d=\StembedEmbeddingDim$, followed by $\ell_2$ normalization.
The presence head has one GELU hidden layer of width \StembedPresenceHidden{}.
The model uses \StembedParametersRoundedM{}M parameters, including \StembedBackboneParametersRoundedM{}M in MuQ through layer \StembedReadout{}.

\section{Experimental Setup}
\label{sec:experiments}

Training uses MoisesDB~\cite{pereira_moisesdb_2023}, whose songs contain 2--20 stems.
Using AudioTree~\cite{Braun_AudioTree}, we measure each stem's loudness on a fixed grid of \qty{5}{\second} windows and treat windows above \qty{-40}{\text{LUFS}} as active.
A usable stem has at least four active windows.
We keep the \SongsTotal{} songs with at least four usable stems and sample only stem subsets for which every required mixture has a window with all constituent stems active.
Recordings connected by a shared artist credit, including featured-artist aliases, form an artist group and belong to the same split.
All training runs share \SongsTrain{}/\SongsValidation{}/\SongsTest{} songs and \TrainArtistGroups{}/\ValidationArtistGroups{}/\TestArtistGroups{} artist groups in the training, validation, and test splits.

Stembed's architecture search includes MuQ depth, decoder size, and embedding dimension.
Checkpoint selection uses validation mean reciprocal rank (MRR), evaluated on 36 validation songs against a gallery of 1755 training and validation stems.
We evaluate every 250 updates and train each selected setting with three seeds.

All trained models use AdamW~\cite{loshchilov_decoupled_2019} ($\beta_1{=}0.9$, $\beta_2{=}0.95$, weight decay 0.1 on matrix parameters, gradients clipped to global norm 10).
Stembed and CIR--MuQ each process \StembedBatchInputs{} five-second inputs per step for \StembedTrainingUpdates{} steps.
The learning rate warms up linearly from 1\% of its peak for 125 updates, then follows cosine decay to $10^{-5}$.
Validation selects Stembed's peak learning rate of $3\times10^{-5}$ from $\{3\times10^{-5},10^{-4}\}$.
Baseline rates are given in \S\ref{ssec:baselines}.

\subsection{Retrieval protocol}
\label{ssec:retrieval}

Although deployment aims to retrieve similar-but-not-exact stems, we measure retrieval with the original stems in the database.
We rank stems by cosine similarity.
A pooled model returns one ranking per mixture, which all of its targets share.
For an $m$-stem query, Stembed returns one ranking for each of the $m$ slots with the highest presence logits.
To emulate a user's choice among slots, an oracle encodes the isolated ground-truth (GT) stems from the query window with the solo path and assigns each target to a distinct selected slot by minimum-cost matching.
The oracle does not change the selected slots or their rankings.
R@$k$ is the fraction of targets whose gallery stem appears in the first $k$ results of the target's ranking.

We generate one fixed evaluation set and gallery, shared by all models and training seeds.
MoisesDB evaluation cycles through the \SongsTest{} test songs in fixed order for \MoisesEvalVisits{} total draws, giving 10 or 11 draws per song.
The standard deviation reported in our tables reflects the variation in training seeds without accounting for the limited number of test songs.
In each draw, we sample four stems that can form audible mixtures and evaluate all four three-stem combinations.
This gives \MoisesEvalMixtures{} mixtures and \MoisesEvalTargets{} target-stem retrievals per checkpoint.
As in CIR~\cite{vaillant_contrastive_2026}, training stems serve as distractors;
the gallery contains one randomly sampled active window per usable stem from all three splits, for \DbSize{} stems in total.
Each gallery excerpt comes from active windows disjoint from that stem's query windows.
For \OverlappingTargetQueries{} of \MoisesEvalTargets{} targets (\OverlappingTargetPercent\%), no such window exists, and the excerpt may overlap a query window.

We report two search protocols, ``All'' and ``Family.''
``All'' searches every database stem.
At rank one, a pooled encoder can retrieve at most one of the $m$ stems in a mixture, limiting All R@1 to $1/m$.
The ``Family'' protocol restricts each search to the target stem's instrument family, following CIR~\cite{vaillant_contrastive_2026}.
We map MoisesDB instrument labels, such as acoustic guitar, to the merged instrument-family taxonomy defined in CIR's released code.

We also evaluate Stembed without the ground-truth stem count $m$.
This ``predicted-count'' variant keeps slots whose presence probability exceeds 50\%, then matches GT stems to distinct slots.
Unmatched targets receive zero recall, and surplus slots remain unused.

\begin{table}[!t]
\centering
\caption{
Three-stem MoisesDB retrieval (percent).
Trained rows report three-seed means and sample standard deviation (SD) for All R@1.
Other SDs are at most \CrossSongNegativesFiveSd{} points.
Frozen rows use one evaluation.
Bold marks column maxima, excluding GT-stem references and ablations.
``All'' searches \DbSize{} stems; ``Family'' searches only the target's family.
}
\label{tab:retrieval}
{\footnotesize\begin{tabular*}{\linewidth}{@{\extracolsep{\fill}}lrrrr@{}}
\toprule
 & \multicolumn{2}{c}{R@1} & \multicolumn{2}{c}{R@5} \\
\cmidrule(lr){2-3}\cmidrule(lr){4-5}
System & All & Family & All & Family \\
\midrule
\multicolumn{5}{l}{\textit{Trained retrieval}} \\
Stembed & $\boldsymbol{56.8\pm0.9}$ & \textbf{58.9} & \textbf{78.1} & 81.2 \\
CIR--MuQ & $23.6\pm0.5$ & 52.4 & 60.7 & \textbf{81.7} \\
SMTI--MuQ & $21.4\pm0.6$ & 23.7 & 39.8 & 44.0 \\
\midrule
\multicolumn{5}{l}{\textit{Frozen encoders}} \\
CIR--AST~\cite{vaillant_contrastive_2026} & 10.7 & 20.1 & 22.3 & 41.3 \\
MuQ~\cite{zhu_muq_2025} & 10.7 & 16.0 & 17.1 & 28.1 \\
MuQ-MuLan~\cite{zhu_muq_2025} & 6.0 & 12.1 & 13.6 & 26.6 \\
FIGMA~\cite{anand_figma_2026} & 10.1 & 20.0 & 20.7 & 39.4 \\
MERT~\cite{li_mert_2024} & 5.4 & 9.4 & 10.8 & 20.2 \\
LAION-CLAP~\cite{wu_large-scale_2023} & 4.3 & 6.7 & 9.7 & 18.6 \\
\midrule
\multicolumn{5}{l}{\textit{Separation and isolated-stem references}} \\
HT-Demucs + Stembed & $53.2\pm0.5$ & 55.9 & 72.3 & 75.8 \\
HT-Demucs + CIR--AST & 34.5 & 36.0 & 55.0 & 57.5 \\
GT stems + Stembed & $75.1\pm0.7$ & 75.5 & 89.1 & 89.6 \\
GT stems + CIR--AST & 64.8 & 65.6 & 81.3 & 81.7 \\
GT stems + frozen MuQ & 52.8 & 53.4 & 65.1 & 66.9 \\
\midrule
\multicolumn{5}{l}{\textit{Stembed ablations}} \\
Predicted count & $56.0\pm1.3$ & 57.8 & 76.3 & 78.7 \\
Slot cosine-triplet & $58.0\pm0.5$ & 60.1 & 79.2 & 81.9 \\
Same-song negatives only & $7.9\pm0.4$ & 9.0 & 18.6 & 21.9 \\
Cross-song negatives only & $45.3\pm1.3$ & 50.1 & 70.3 & 77.8 \\
Shared window & $57.0\pm0.7$ & 58.9 & 75.0 & 77.3 \\
Random cross-song mixtures & $57.5\pm0.9$ & 59.4 & 76.7 & 78.9 \\
\bottomrule
\end{tabular*}
}
\end{table}

\noindent\textbf{Slakh2100 evaluation.}
Using the same protocol, we evaluate the checkpoints on Slakh2100~\cite{manilow_cutting_2019}.
We use \SlakhSongs{} multitrack test songs with three draws per song.
As in the MoisesDB evaluation, we use all four three-stem combinations that can be made from four stems.
This results in a total of \SlakhEvalMixtures{} three-stem mixtures.
The gallery contains one active excerpt per usable stem from the test recordings only, giving \SlakhDbSize{} stems.
We map Slakh's General MIDI classes to CIR's families using the mapping in our released code.
Unmapped classes are excluded from both queries and the gallery.

\subsection{Baselines}
\label{ssec:baselines}

\noindent\textbf{CIR--MuQ.}
This baseline applies CIR's full-triplet objective~\cite{vaillant_contrastive_2026} to mean-pooled MuQ embeddings.
Each mixture-stem positive pair forms triplets in both anchor directions, using every other embedding as a negative, including sibling stems.
The objective uses Euclidean distance with margin 1.0.
Each group contributes three isolated stems and their three-stem mixture.
Groups use distinct stems within a batch to avoid treating a stem as its own negative.
Validation independently selects the first \CirMuqReadout{} MuQ layers (\CirMuqParametersRoundedM{}M parameters), versus Stembed's \StembedReadout{} layers.
The selected peak learning rate is $10^{-4}$ from $\{3,5,10,20,40\}\times10^{-5}$.

\noindent\textbf{SMTI--MuQ.}
Following SMTI~\cite{kim_show_2023}, this baseline trains a mixture encoder to predict embeddings from a frozen solo encoder.
The teacher averages MuQ's final-layer frames into \SmtiEmbeddingDim{}-dimensional embeddings.
The student uses Stembed's backbone and decoder with a \SmtiEmbeddingDim{}-dimensional output head.
Minimum-cost matching assigns student slots to teacher embeddings, and training minimizes their mean cosine distance plus Stembed's presence loss.
Both embeddings are normalized before matching.
The teacher also encodes the gallery.
SMTI--MuQ trains for \SmtiTrainingUpdates{} updates using the same 32 graphs per batch as Stembed.
The student encodes the six mixtures per graph, giving \SmtiMixtureInputs{} inputs per update.
The teacher encodes each mixture's isolated constituents at the corresponding time window, giving $32\times(2\times2+4\times3)=\SmtiTeacherInputs{}$ reference inputs.
Validation MRR selects peak rate $10^{-4}$.

\noindent\textbf{Frozen encoders.}
We compare five general-purpose encoders~\cite{zhu_muq_2025,anand_figma_2026,li_mert_2024,wu_large-scale_2023} and the released CIR--AST full-triplet checkpoint~\cite{vaillant_contrastive_2026}, which fine-tuned a \qty{16}{\kilo\hertz} AST~\cite{gong_ast_2021} on NSynth notes~\cite{engel_neural_2017} and synthesizer renders.
MuQ uses mean-pooled final-layer frames, and FIGMA uses its 512-dimensional projected audio embedding.

\noindent\textbf{Separation and isolated-stem references.}
HT-Demucs~\cite{rouard_hybrid_2023}, using the \texttt{htdemucs\_6s} checkpoint, separates \qty{44.1}{\kilo\hertz} stereo mixtures into six outputs.
For separation, we construct \qty{44.1}{\kilo\hertz} stereo mixtures from the original stems using the same stem identities and time windows as the retrieval queries.
We downmix the separated outputs to mono and resample them for Stembed or CIR--AST.
Each encoder embeds these outputs, the isolated GT references, and the gallery.
Oracle matching assigns targets to separator outputs as in \S\ref{ssec:retrieval}.
The GT-stem rows bypass separation and query the gallery with the original isolated excerpts.
``GT stems + frozen MuQ'' uses SMTI--MuQ's teacher.

\section{Results}

\label{sec:results}

\noindent\textbf{Stembed and trained baselines.}
Stembed reaches \StembedOne\% All R@1, compared with \CirMuqOne\% for CIR--MuQ (Table~\ref{tab:retrieval}).
A pooled encoder returns one ranking per mixture, so only one of three target stems can rank first, limiting its All R@1 to 33.3\%.
With family guidance, Stembed has higher R@1 (\StembedFamilyOne\% versus \CirMuqFamilyOne\%), while CIR--MuQ has slightly higher R@5 (\CirMuqFamilyFive\% versus \StembedFamilyFive\%).
SMTI--MuQ reaches \SmtiOne\% All R@1.
Querying with the frozen teacher's embeddings of isolated stems reaches \GtFrozenMuqOne\% All R@1, compared with \GtStembedOne\% for Stembed's solo embeddings of the same excerpts.
Both imitation error and the teacher's representation limit SMTI--MuQ.

\noindent\textbf{Frozen encoders.}
Frozen encoders reach at most \CirOne\% All R@1 and remain below CIR--MuQ even with family filtering.
MuQ-MuLan, MERT, and LAION-CLAP reach at most \MuqMulanOne\% All R@1.

\noindent\textbf{Separation and isolated-stem references.}
HT-Demucs + Stembed reaches \DemucsStembedOne\% All R@1, below direct Stembed retrieval's \StembedOne\%, despite allowing the oracle to choose among six separator outputs rather than three selected slots.
HT-Demucs + CIR--AST reaches \DemucsCirOne\%, far above pooled CIR--AST's \CirOne\%.
Queries from isolated stems reach \GtStembedOne\% with Stembed, so extracting a target representation from a mixture costs about \num[evaluate-expression,round-mode=places,round-precision=0]{\GtStembedOne-\StembedOne} points.
The remaining misses reflect the difficulty of matching different excerpts of the same stem.

\noindent\textbf{Same-artist distractors.}
An exact-stem metric penalizes retrieving a similar instrument from another song by the same artist.
Therefore, we evaluate on MoisesDB again by removing gallery stems from other songs in the query's artist group.
Removing these distractors raises All R@1 by \num[evaluate-expression]{\FilteredStembedOne-\StembedOne} points for Stembed, \num[evaluate-expression]{\FilteredPredictedCountOne-\PredictedCountOne} with predicted counts, \num[evaluate-expression]{\FilteredCirMuqOne-\CirMuqOne} for CIR--MuQ, and \num[evaluate-expression]{\FilteredSmtiOne-\SmtiOne} for SMTI--MuQ.
Stembed's larger gain indicates that more targets were outranked only by stems from other songs by the same artist.

\noindent\textbf{Slakh2100 evaluation.}
On Slakh2100 (Table~\ref{tab:slakh}), both count-supplied and predicted-count Stembed exceed CIR--MuQ and SMTI--MuQ in All R@1 and All R@5.
With family guidance, count-supplied Stembed and CIR--MuQ have close R@1 values.
CIR--MuQ has higher Family R@5.

\subsection{Ablations}

\noindent\textbf{Predicted count.}
Thresholding presence at 50\% without supplying the stem count gives $\PredictedCountOne\pm\PredictedCountOneSd\%$ All R@1 and \PredictedCountFive\% R@5 (Table~\ref{tab:retrieval}), close to count-supplied retrieval.
On the test draws, exact-count accuracy is \CountTwoAccuracy\%, \CountThreeAccuracy\%, and \CountFourAccuracy\% for two-, three-, and four-stem mixtures.
The detector overcounts \CountTwoOver\% of two-stem mixtures and undercounts \CountFourUnder\% of four-stem mixtures, suggesting a bias toward three stems.
Training includes two- and three-stem mixtures but no four-stem mixtures, which may explain the undercounting.
The remaining ablations use the supplied count.

\noindent\textbf{Slot-based cosine-triplet control.}
\label{ssec:objective-controls}
Replacing InfoNCE with cosine-triplet training over all positive--negative combinations, including zero-loss triplets, gives $\SlotTripletOne\pm\SlotTripletOneSd\%$ All R@1, compared with $\StembedOne\pm\StembedOneSd\%$ for InfoNCE.
Validation MRR selects margin $\num[round-mode=figures,round-precision=1]{\TripletMargin}$ and weight $\num[round-mode=places,round-precision=0]{\TripletWeight}$ from $\{0.1, 0.2, 0.4, 0.8\}\times\{0.25, 1, 4, 16\}$, with presence weight 0.1.
Its higher mean recall suggests that InfoNCE is not essential to Stembed's gains.
CIR--MuQ also uses triplets but pools each mixture into one embedding.
Relative to CIR--MuQ, the slot-based cosine-triplet raises All R@5 from \CirMuqFive\% to \SlotTripletFive\% and Family R@1 from \CirMuqFamilyOne\% to \SlotTripletFamilyOne\%.
This suggests slot-based training benefits retrieval more than the choice of contrastive loss does.
However, the two models also differ in distance function, margin, and MuQ depth.

\noindent\textbf{Negative pairs.}
Each control drops one negative category from Eq.~\eqref{eq:contrastive}'s denominator and keeps all positives: ``same-song negatives only'' drops cross-song negatives, and ``cross-song negatives only'' drops same-song negatives.
Same-song negatives alone reach $\SameSongNegativesOne\pm\SameSongNegativesOneSd\%$ All R@1, while cross-song negatives alone achieve $\CrossSongNegativesOne\pm\CrossSongNegativesOneSd\%$.
Both fall below the full model, particularly without cross-song negatives, which support discrimination across the library.

\noindent\textbf{Independent time windows.}
Sharing a window across all ten nodes changes mean All R@1 from \StembedOne\% to \SharedWindowOne\% but reduces R@5 from \StembedFive\% to \SharedWindowFive\%.
Independent windows help R@5 in this comparison, without improving R@1.

\noindent\textbf{Random cross-song mixtures.}
Sampling each graph's stems across training songs, with independent windows and unchanged audibility requirements, changes mean All R@1/R@5 from \StembedOne\%/\StembedFive\% to \CrossSongMixturesOne\%/\CrossSongMixturesFive\%.
This suggests that temporally aligned stems from the same song are not necessary for constructing effective training mixtures in our framework.

\begin{table}[!t]
\centering
\caption{
Three-stem retrieval on Slakh2100 using models selected on MoisesDB.
Values are mean $\pm$ sample SD (\%) across three training seeds.
The gallery contains \SlakhDbSize{} stems, with search protocols defined in Table~\ref{tab:retrieval}.
Bold marks the best mean in each column.
}
\label{tab:slakh}
{\footnotesize\setlength{\tabcolsep}{2pt}\begin{tabular*}{\linewidth}{@{\extracolsep{\fill}}lrrrr@{}}
\toprule
 & \multicolumn{2}{c}{R@1} & \multicolumn{2}{c}{R@5} \\
\cmidrule(lr){2-3}\cmidrule(lr){4-5}
System & All & Family & All & Family \\
\midrule
Stembed & $\boldsymbol{38.1\pm1.1}$ & $\boldsymbol{43.2\pm1.1}$ & $\boldsymbol{59.0\pm1.5}$ & $68.0\pm1.2$ \\
Predicted count & $37.5\pm0.1$ & $41.4\pm0.4$ & $56.7\pm0.7$ & $62.9\pm0.4$ \\
CIR--MuQ & $20.2\pm0.3$ & $42.6\pm0.5$ & $45.1\pm0.3$ & $\boldsymbol{70.0\pm0.7}$ \\
SMTI--MuQ & $15.8\pm0.7$ & $19.3\pm0.5$ & $29.4\pm0.5$ & $36.8\pm0.6$ \\
\bottomrule
\end{tabular*}
}
\end{table}

\section{Conclusion}
\label{sec:conclusion}

Stembed encodes a mixture as several stem-specific embeddings by combining slot decoding, minimum-cost matching, and contrastive learning.
On MoisesDB, it reaches recall comparable to family-guided CIR--MuQ without instrument-family labels.

Future work could evaluate perceptual similarity when the target is absent from the database, adapting Stem-JEPA's~\cite{riou_stem-jepa_2024} listening study to assess similarity rather than compatibility.
We plan to train on denser mixtures and increase the number of slots as needed.
Slot embeddings could also condition generative models or support systems that reconstruct mixtures through retrieval and resynthesis.

\clearpage
\section{Acknowledgments}
This work was supported in part by the National Science Foundation under award numbers IIS-2523648 and IIS-2523649.

\bibliographystyle{IEEEbib}
\bibliography{main}

\end{document}